\documentclass[prd,amsmath,amssymb,superscriptaddress,twocolumn,10pt]{revtex4}%superscriptaddress,showpacs,

\pdfoutput=1

\usepackage{graphicx}
\usepackage{dcolumn}
\usepackage{bm}
\usepackage{amssymb}
\usepackage{latexsym}
\usepackage{booktabs}
\usepackage{amsmath}
\usepackage{multirow}
\usepackage[colorlinks=true, citecolor=blue]{hyperref}

\newcommand{\Mpc}{\mathrm{~km~s^{-1}~Mpc^{-1}}}

\def\({\left(}
\def\){\right)}
\def\[{\left[}
\def\]{\right]}

\begin{document}
\title{Measurements of the Hubble constant and cosmic curvature with
quasars: ultra-compact radio structure and strong gravitational
lensing}

\author{Jing-Zhao Qi}
\affiliation{Department of Physics, College of Sciences, Northeastern
University, Shenyang 110819, China}

\author{Jia-Wei Zhao}
\affiliation{Department of Physics, College of Sciences, Northeastern
University, Shenyang 110819, China}

\author{Shuo Cao}
\email{caoshuo@bnu.edu.cn}
\affiliation{Department of Astronomy, Beijing Normal University,
Beijing, 100875, China}

\author{Marek Biesiada}
\affiliation{National Centre for Nuclear Research, Pasteura 7,
02-093 Warsaw, Poland}

\author{Yuting Liu}
\affiliation{Department of Astronomy, Beijing Normal University,
Beijing, 100875, China}

\begin{abstract}
Although the Hubble constant $H_0$ and spatial curvature
$\Omega_{K}$ have been measured with very high precision, they still
suffer from some tensions.
%most of
%these measurements were still suffering from the well-known Hubble tension and cosmic curvature tension.
In this paper, we propose an improved method to combine the
observations of ultra-compact structure in radio quasars and strong
gravitational lensing with quasars acting as background sources to
determine $H_0$ and $\Omega_{K}$ simultaneously. By applying the
distance sum rule to the time-delay measurements of 7 strong lensing
systems and 120 intermediate-luminosity quasars calibrated as
standard rulers, we obtain stringent constraints on the Hubble
constant ($H_0=78.3\pm2.9 \Mpc$) and the cosmic curvature
($\Omega_K=0.49\pm0.24$). On the one hand, in the framework of a
flat universe, the measured Hubble constant ($H_0=73.6^{+1.8}_{-1.6}
\Mpc$) is strongly consistent with that derived from the local
distance ladder, with a precision of 2\%. On the other hand, if we
use the local $H_0$ measurement as a prior, our results are
marginally compatible with zero spatial curvature
($\Omega_K=0.23^{+0.15}_{-0.17}$) and there is no significant
deviation from a flat universe. Finally, we also evaluate whether
strongly lensed quasars would produce robust constraints on $H_0$
and $\Omega_{K}$ in the non-flat and flat $\Lambda$CDM model, if the
compact radio structure measurements are available from VLBI
observations.
\end{abstract}

%\keywords{Quasars (1319); Strong gravitational lensing (1643);
%Cosmological parameters (339); Radio sources (1358)}

\maketitle

\section{Introduction}

The Hubble constant $H_0$, which sets the present expansion rate of
the Universe, has been one of the most important cosmological
parameters one attempted to measure. Although $H_0$ has been
measured with very high precision, these measurements are currently
in tension with each other.
In the framework of the flat $\Lambda$ cold dark matter
($\Lambda$CDM) cosmological model which is proven to be well
consistent with various observations
\cite{cao2010testing,cao2012SL,cao2014cosmic,Cao:2015qja}, the
\textit{Planck} satellite provides a stringent constraint on the
Hubble constant as $H_0=67.4\pm0.5 \Mpc$, based on the anisotropies
of the Cosmic Microwave Background Radiation (CMBR)
\cite{Aghanim:2018eyx}. The other independent estimates of $H_0$
have been obtained by local type Ia supernovae (SNe Ia) calibrated
via the distance ladder, with the value of $H_0=74.03\pm 1.42 \Mpc$
recently released by SH0ES (SNe, H0, for the Equation of State of
dark energy) collaboration \cite{Riess:2019cxk}. However, such
significant 4.4 $\sigma$ tension can not be solely attributed to
systematics errors
\cite{Riess:2019cxk,DiValentino:2018zjj,Follin:2017ljs}. Therefore,
any other independent way to measure the Hubble constant is of
utmost importance.
%several geometrical methods should be
%taken into account to provide independent measurements of the Hubble constant.

As early as in 1964, it has been proposed \cite{Refsdal:1964nw}
that observations of time delays from the gravitationally lensed
supernovae could be used to determinate $H_0$ . Even though, the
lensed supernovae have eventually been detected
\cite{Kelly:2014mwa,Goobar:2016uuf}, the sample is far too small to
contribute in solving of $H_0$ tension. In fact, more promising
candidates in this context are quasars. Due to the brightness and
variable nature of quasars, some recent works focused on the lensed
quasars to measure the so-called "time-delay distance" ($D_{\Delta
t}$), a combination of three angular diameter distances between the
observer, lens, and source sensitive to $H_0$. More recently, the
H0LiCOW ($H_0$ Lenses in COSMOGRAIL's Wellspring) collaboration
showed that real observations of time delays for six lensed quasars
can be used to obtain the Hubble constant $H_0=73.3^{+1.7}_{-1.8}
\Mpc$ in the spatially flat $\Lambda$CDM model \cite{Wong:2019kwg}.
The result is in perfect agreement with the local measurements of
$H_0$ from the SH0ES collaboration. However, the drawback of this
method is that the fits on $H_0$ are strongly model dependent, i.e.,
the value of $H_0$ would shift to $81.6^{+4.9}_{-5.3} \Mpc$ when the
equation of state (EoS) parameter of dark energy is treated as a
free parameter.

In fact, the $H_0$ tension suggests the possibility that there could
be an inconsistency between the early-universe and late-universe in
modern cosmological theories. Moreover, the recent studies
\cite{DiValentino:2019qzk,DiValentino:2020hov,2019arXiv190809139H}
of the spatial curvature parameter $\Omega_K$ also highlighted
inconsistency. A higher lensing amplitude, $A_{\rm{lens}}$ in the
CMB power spectra has been confirmed by the recent Planck 2018
results \cite{Aghanim:2018eyx}. Closed Universe can provide a
physical explanation for this effect, resulting with
$\Omega_K=-0.044^{+0.018}_{-0.015}$ constrained from the combination
of the \textit{Planck} temperature and polarization power spectra
\cite{Aghanim:2018eyx}. However, combining the \textit{Planck}
lensing with low-redshift baryon acoustic oscillations (BAO),
changes this constraint to $\Omega_K = 0.0007\pm{0.0019}$ precisely
fitting the flat Universe as it was expected by the inflation
theory. In further analysis, Di Valentino et al. \cite{DiValentino:2019qzk} showed that
under a closed Universe preferred by \textit{Planck}, higher than
generally estimated discordances arise for most of the local
cosmological observables. However, it should be stressed here that
this method also makes a strong assumption based on some specific
dark energy model, i.e., the non-flat $\Lambda$CDM model. To better
understand the discrepancy between the Hubble constant and cosmic
curvature measured locally and the value inferred from the
\textit{Planck} survey, fully model-independent measurements of
$H_0$ and $\Omega_K$ are still required. For the discussions about $H_0$ tension, we
refer the reader the following works \cite{guo2019can,vagnozzi2020new,zhang2019gravitational,qi2020new,vattis2019dark,zhang2014neutrinos,zhao2017search,guo2017constraints}

In particular, the distance sum rule \cite{Rasanen:2014mca}
provides an effective approach to determine the spatial curvature
and the Hubble constant but without adopting any particular model.
In the framework of such a theoretical framework, great efforts have
been made in the recent studies to set limits on the cosmic
curvature
\cite{Qi:2018atg,Wang:2019yob,Xia:2016dgk,Li:2018hyr,Zhou:2019vou},
based on the precise measurements of the source-lens/observer
distance ratio in strong gravitational lensing (SGL) systems
\cite{Cao:2015qja,Chen:2018jcf} and luminosity distances in other
different distance indicators \cite{Qi:2018aio,Liao:2019hfl}.
However, such methodology is weakly dependent on the Hubble
constant, which enters not through a distance measure directly, but
rather through a distance ratio \cite{cao2012constraints}. Based on
the precise measurements of time delays between multiple images, the
first attempt to determine the spatial curvature and Hubble constant
with SN Ia luminosity distances was presented in
\cite{Collett:2019hrr}. This approach was then extended by using
another luminosity distance probe, i.e. the nonlinear relation
between the ultraviolet (UV) and X-ray monochromatic luminosities
satisfied by high-redshift quasars \cite{Wei:2020suh}.

However, one should remember that for the implementation of the
distance sum rule, it would be beneficial to use distance probes
covering higher redshifts thus taking advantage of a larger sample
of SGL systems. More importantly, the cosmological application of
time delay measurements requires good knowledge of three angular
diameter distances of the lensed quasar systems (from observer to
lens, from observer to source, and from lens to source). Therefore,
a new promising window of opportunity would open up if we could
extend the $H_0$ and $\Omega_K$ measurements by using a new, deeper
astronomical probes acting as standard rulers in a redshift range
well consistent with the lensed quasars \cite{Qi:2018aio}.

In this paper, we will further study $H_0$ and $\Omega_K$ using the
angular size of compact structure in radio quasars as standard
rulers and the time delay from lensed quasars. By virtue of the
distance sum rule, a cosmological model-independent analysis becomes
possible. In order to investigate the influence of the cosmological
model on the constraints of $H_0$ and $\Omega_K$, we will also
perform a comparative analysis of the non-flat and flat $\Lambda$CDM
models.

%others \cite{Yu:2016gmd,Wang:2020dbt}

\section{Methodology and observations}

In the homogeneous and isotropic universe, the FLRW metric is
applied to describe its spacetime:
\begin{equation} \label{flrw}
ds^2= c^2dt^2 - \frac{a(t)^2}{1-Kr^2}dr^2 - a(t)^2r^2d\Omega^2,
\end{equation}
where $a(t)$ represents the scale factor and $c$ is the speed of
light. Note that the cosmic curvature parameter $\Omega_K$ is
determined by the dimensionless curvature $K$ and the Hubble
constant $H_0$ as $\Omega_K=-K c^2 /a_0^2 H_0^2$. In this analysis
we focus on a specific strong lensing system with the background
quasar (at redshift $z_s$) as the source and the early-type galaxy
(at redshift $z_l$) acting as a lens . By introducing dimensionless
comoving distances $d_{ls}\equiv d(z_l,z_s)$, $d_l\equiv d(0,z_l)$
and $d_s\equiv d(0,z_s)$, these three types of distances are
connected as
\begin{equation} \label{smr}
\frac{d_{ls}}{d_s}=\sqrt{1+\Omega_Kd_l^2}-\frac{d_l}{d_s}\sqrt{1+\Omega_K
d_s^2}.
\end{equation}
by virtue of the well-known distance sum rule in non-flat FLRW
models. The original idea of cosmological application of the
distance sum rule in general, and with respect to gravitational
lensing data in particular, can be traced back to the paper of Ref.
\cite{Rasanen:2014mca}, which has been extensively discussed in
more recent papers focused on testing the spatial curvature of the
Universe \cite{Qi:2018aio} and thus the validity of the FLRW metric
\cite{Qi:2018atg,cao2019direct}. Furthermore, one is able to
rewrite Eq.~(\ref{smr}) as
\begin{equation} \label{timedelay1}
\frac{d_{l}d_{s}}{d_{ls}}=\frac{1}{\sqrt{1/d_l^2+\Omega_K}-\sqrt{1/d_s^2+\Omega_K}}.
\end{equation}
Let us note that the dimensionless comoving distances $d$ are
related to the (dimensioned) comoving distances $D$ as $d=H_0D/c$.

\textit{Time-delay measurements from lensed quasars.---} In the
framework of a strong lensing system with the background quasar as
the source and the early-type galaxy acting as a lens, one of the
typical feature is that time delays between lensed images
($\mathbf{\theta}_i$ and $\mathbf{\theta}_j$) are dependent
on the time-delay distance ($D_{\mathrm{\Delta
t}}\equiv(1+z_l)\frac{D^A_{\mathrm{l}}
D^A_{\mathrm{s}}}{D^A_{\mathrm{ls}}}$) and the gravitational
potential of the lensing galaxy as
\begin{equation}
\Delta t_{i,j} = \frac{D_{\mathrm{\Delta t}}}{c}\Delta \phi_{i,j},
\label{relation}
\end{equation}
where
$\Delta\phi_{i,j}=[(\mathbf{\theta}_i-\mathbf{\beta})^2/2-\psi(\mathbf{\theta}_i)-(\mathbf{\theta}_j-\mathbf{\beta})^2/2+\psi(\mathbf{\theta}_j)]$
is the Fermat potential difference determined by the two-dimensional
lensing potential $\psi$ and the source position
$\mathbf{\beta}$. Therefore, the time-delay distance, i.e., the
combination of three angular angular diameter distances, could be
rewritten as
\begin{equation}\label{Ddt}
D_{\mathrm{\Delta t}}=\frac{c \Delta t_{i,j}}{\Delta
\phi_{i,j}}=\frac{c}{H_0}\frac{d_{l}d_{s}}{d_{ls}}.
\end{equation}
As can be clearly seen from Eq.~(\ref{timedelay1}) and (\ref{Ddt}),
if the other two dimensionless comoving distances $d_l$ and $d_s$
can be measured, then the value of $H_0$ and $\Omega_K$ could be
determined directly from the time-delay measurements of $\Delta t$
and well-reconstructed lens potential of $\Delta\phi$, without
involving any specific cosmological model.

For the source of available quasar-galaxy lensing systems, we use
the latest sample of strong-lensing systems with time delay
observations, recently released by the H0LiCOW collaboration and the
STRIDES collaboration with precise time-delay distance measurement
for each lensing system \cite{Wong:2019kwg,shajib2019strides}. The
seven lenses with measured time delays consist of the following
systems: B1608+656 \cite{Suyu10,Jee19}, RXJ1131-1231
\cite{Suyu13,Suyu14,Chen19}, HE 0435-1223 \cite{Wong17,Chen19},
1206+4332 \cite{Birrer19}, WFI2033-4723 \cite{Rusu20}, PG 1115+080
\cite{Chen19}, and DES J0408-5354 \cite{shajib2019strides}, while
the source redshift covers the range of $0.654<z_s<2.375$. The
relevant information necessary to perform statistical analysis,
including the redshifts of both lens and source, as well as the
posterior distributions of the time-delay distances in the form of
Monte Carlo Markov chains (MCMC) are summarized in Ref.
\cite{Wong:2019kwg}. We remark here that a kernel density estimator
was used to compute the posterior distributions of five lenses
(${\cal L}_{\Delta t}$), while the $D_{\Delta t}$ likelihood
function for the lens B1608+656 was given as a skewed log-normal
distribution, due to the absence of blind analysis with respect to
the cosmological quantities of interest.

\textit{Distance calibration from unlensed radio quasars.---} From
the observational point of view, to obtain model-independent
measurements of the distance $d(z)$, one can turn to the objects of
known (or standardizable) comoving size acting as standard rulers.
In this paper, with the aim of deriving the distances to the lens
and the source corresponding to their redshifts, we focus on the
angular size of the compact structure in radio quasars, based on the
very-long-baseline interferometry (VLBI) observations
\cite{Cao:2016dgw}. After refining their selection technique and
redshift measurements, Cao et al. \cite{Cao:2017ivt,Cao:2017abj} collected a
final sample of 120 intermediate-luminosity radio quasars with
reliable measurements of the angular size of the compact structure.
Such recently compiled milliarcsecond compact radio-quasar catalog
covering the redshift range $0.46<z<2.76$ will be used for the
analysis performed in this paper. The angular size of the compact
structure in radio quasars, $\theta(z)$, can be expressed in term of
the angular diameter distance and the linear size of the standard
ruler as
\begin{equation}
\theta(z)=\frac{l_m}{D_A(z)}.
\end{equation}
The angular diameter distance $D_A(z)$ can then be extracted from
the angular size of the compact radio quasars $\theta(z)$, combined
with the linear size of the standard ruler calibrated to
$l_m=11.0\pm0.4$ pc through a new cosmology-independent technique
(the well-measured angular diameter distances from the BAO)
\cite{cao2019milliarcsecond}. Now for each lensing system, one can
obtain the dimensionless distances to the lenses $d_l$ and to the
sources $d_s$ from angular diameter distances to the quasars as:
$d_l=H_0/c(1+z_l)D_A(z_l)$ and $d_s=H_0/c(1+z_s)D_A(z_s)$,
respectively. Considering uncertainties of the angular size
measurements, instead of matching objects by redshift, we decided to
use quasars for reconstructing the dimensionless co-moving distance
function $d(z)$ parameterized by a third-order polynomial
\begin{equation}
d(z)=z+a_1z^2+a_2z^3,
\end{equation}
with the initial conditions of $d(0)=0$ and $d^{'}(0)=1$. The radio
quasar sample is sufficient to reconstruct the profile of $d(z)$ up
to the redshifts $z\sim 3$, without confining ourselves to any
specific cosmology \cite{Qi:2018aio}. Note that the two
coefficients ($a_1$, $a_2$) in a third-order polynomial will be
optimized along with the Hubble constant ($H_0$) and cosmic
curvature ($\Omega_K$). For the radio quasar sample, the posterior
likelihood ${\cal L}_{\rm{QSO}} \sim \exp{(- \chi^2_{\rm{QSO}} /
2)}$ is constructed through the following formula of
\begin{equation}
\label{eq:chi2} \chi^{2}_{\rm{QSO}}=
  \sum_{i}^{120}{\frac{\left[\theta(z_{i}; d(z_i))
     - \theta_{oi}\right]^{2}}{\sigma_{i}^{2}}},
\end{equation}
where $\theta_{oi}$ is the observed angular size for the $ith$
quasar with uncertainty of $\sigma_{i}$. Following the error
strategy proposed in Ref. \cite{Cao:2017ivt}, an additional 10\%
systematical uncertainty in the observed angular sizes is also
assumed in computing ${\cal L}_{\rm{QSO}}$, in order to account for
the intrinsic spread in linear sizes \cite{Cao:2016dgw}.

\begin{figure}
\includegraphics[scale=0.4]{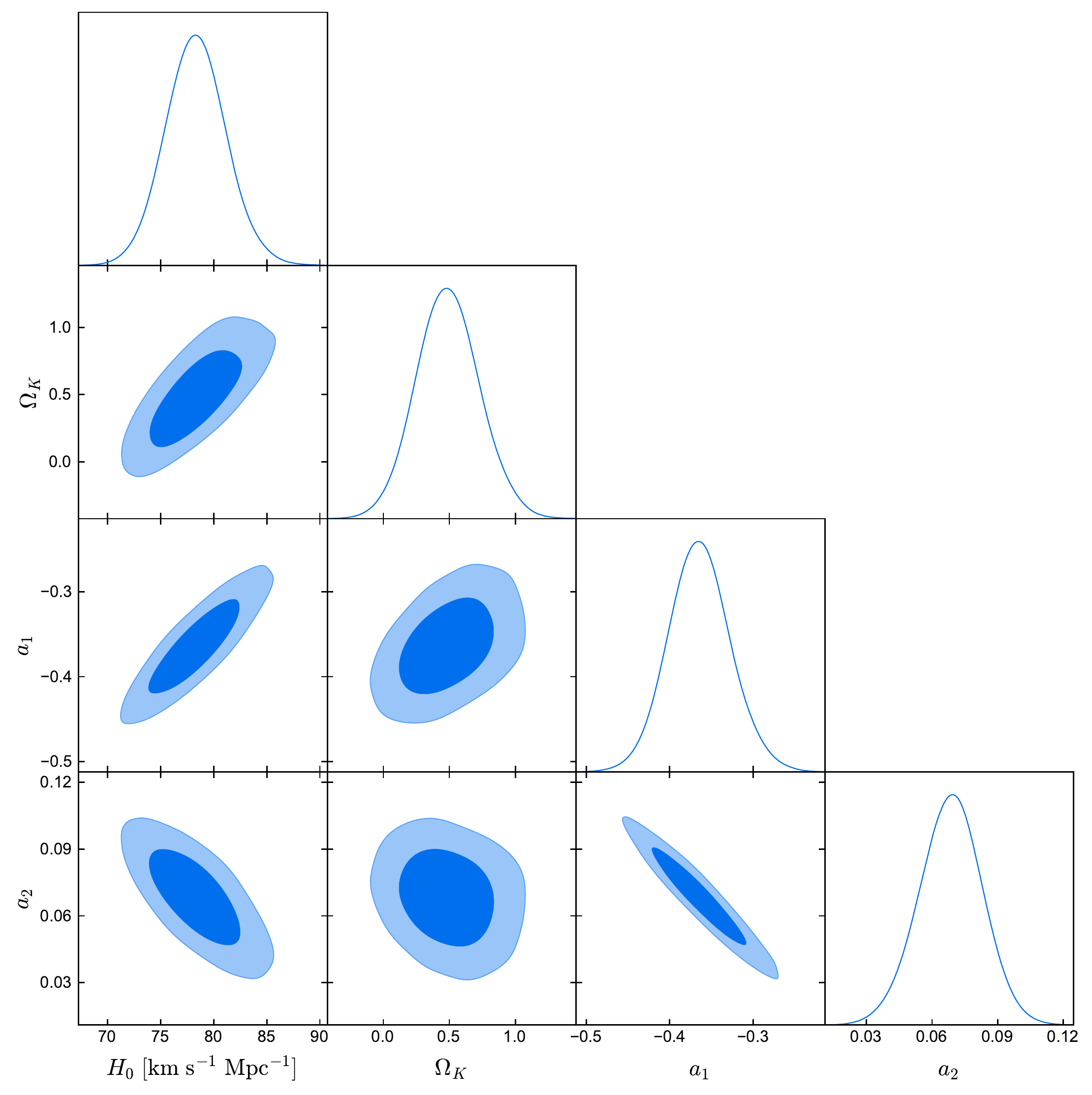}
\caption{Constraints on the parameters $H_0$, $\Omega_K$, $a_1$ and
$a_2$ with strong lensing systems and radio quasars, in the
framework of distance sum rule.\label{Fig1}}
\end{figure}

\begin{figure}
\centering
\includegraphics[scale=0.35]{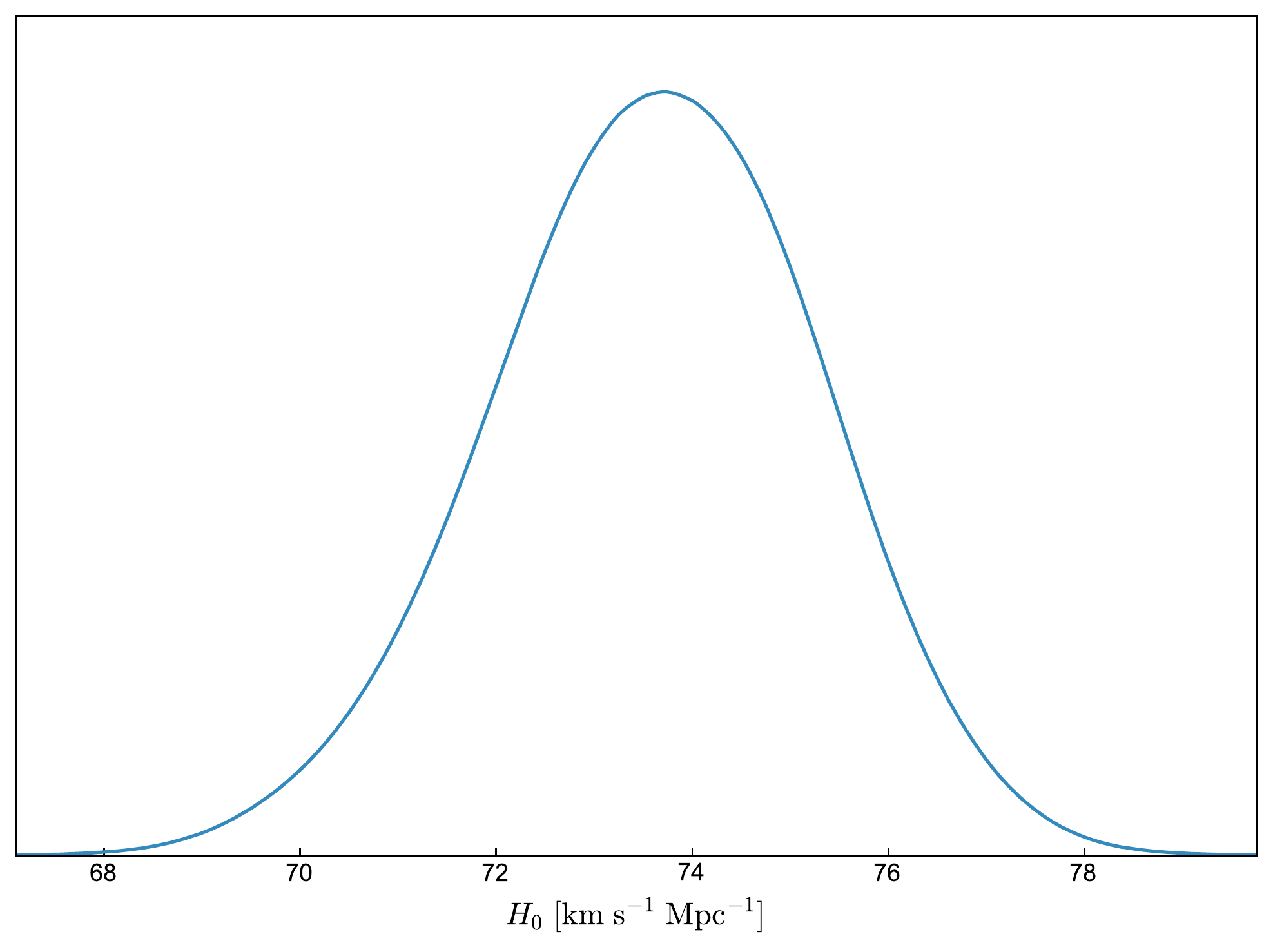}
\includegraphics[scale=0.35]{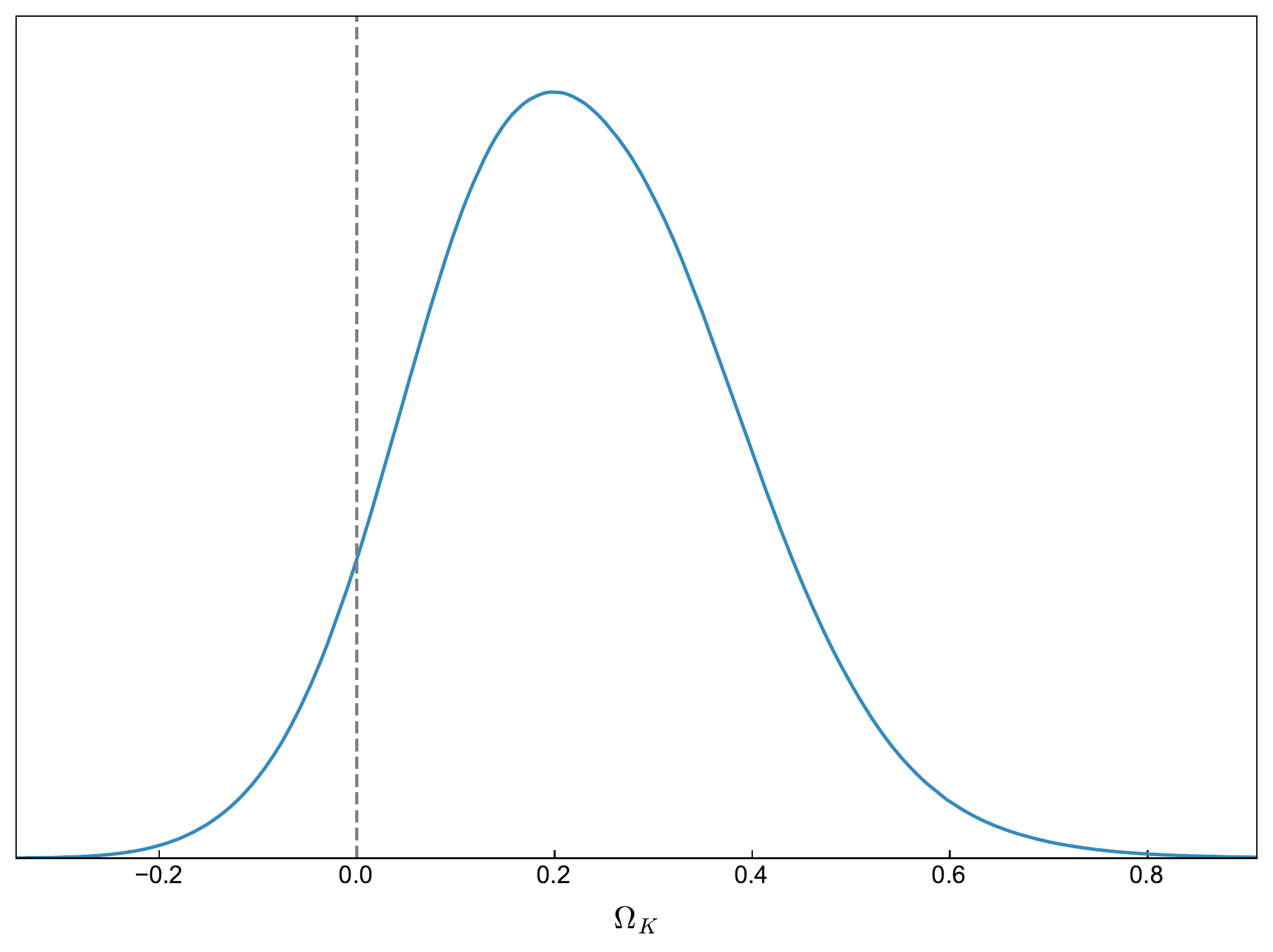}
\caption{Upper panel: Posterior probability distribution of $H_0$
with fixed cosmic curvature ($\Omega_K=0$). Lower panel: Posterior
probability distribution of $\Omega_K$ with fixed Hubble constant
($H_0=74.03 \Mpc$ ).\label{Fig2}}
\end{figure}

\begin{table}
\centering
\caption{Results for $H_0$,
$\Omega_K$ and the parameters of polynomial $a_1$, $a_2$ in the
framework of distance sum rule. \label{tabresult}} 
\begin{tabular}{ccccc}
 & $H_0[\Mpc]$ & $\Omega_K$ &$a_1$ & $a_2$  \\
\hline
& $78.3\pm2.9$ & $0.49\pm0.24$ & $-0.364\pm0.037$ & $0.068\pm0.014$ \\
\hline
& $73.6^{+1.8}_{-1.6}$ & $0$ (fixed) & $-0.404\pm0.030$ & $0.077\pm0.013$ \\
\hline
& $74.03$ (fixed) & $0.23^{+0.15}_{-0.17}$ & $-0.412\pm0.018$ & $0.084\pm0.010$ \\
\hline
\end{tabular}
\end{table}

\begin{figure}
\centering
\includegraphics[scale=0.4]{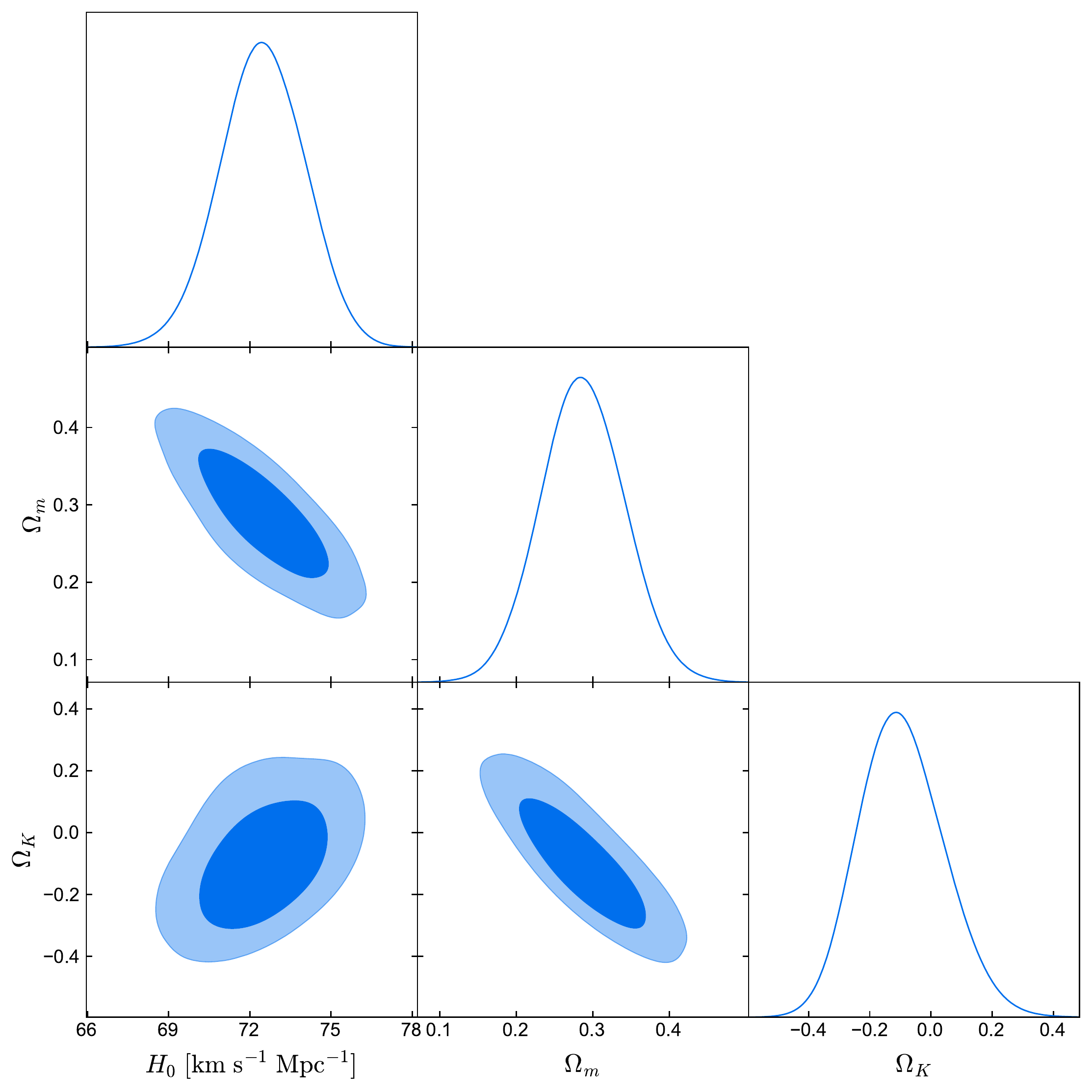}
\caption{1D and 2D marginalized probability distributions for the
parameters $H_0$, $\Omega_m$ and $\Omega_K$ with strong lensing
systems and radio quasars, in the framework of non-flat $\Lambda$CDM
model. \label{OLCDM}}
\end{figure}

\begin{figure}
\centering
\includegraphics[scale=0.35]{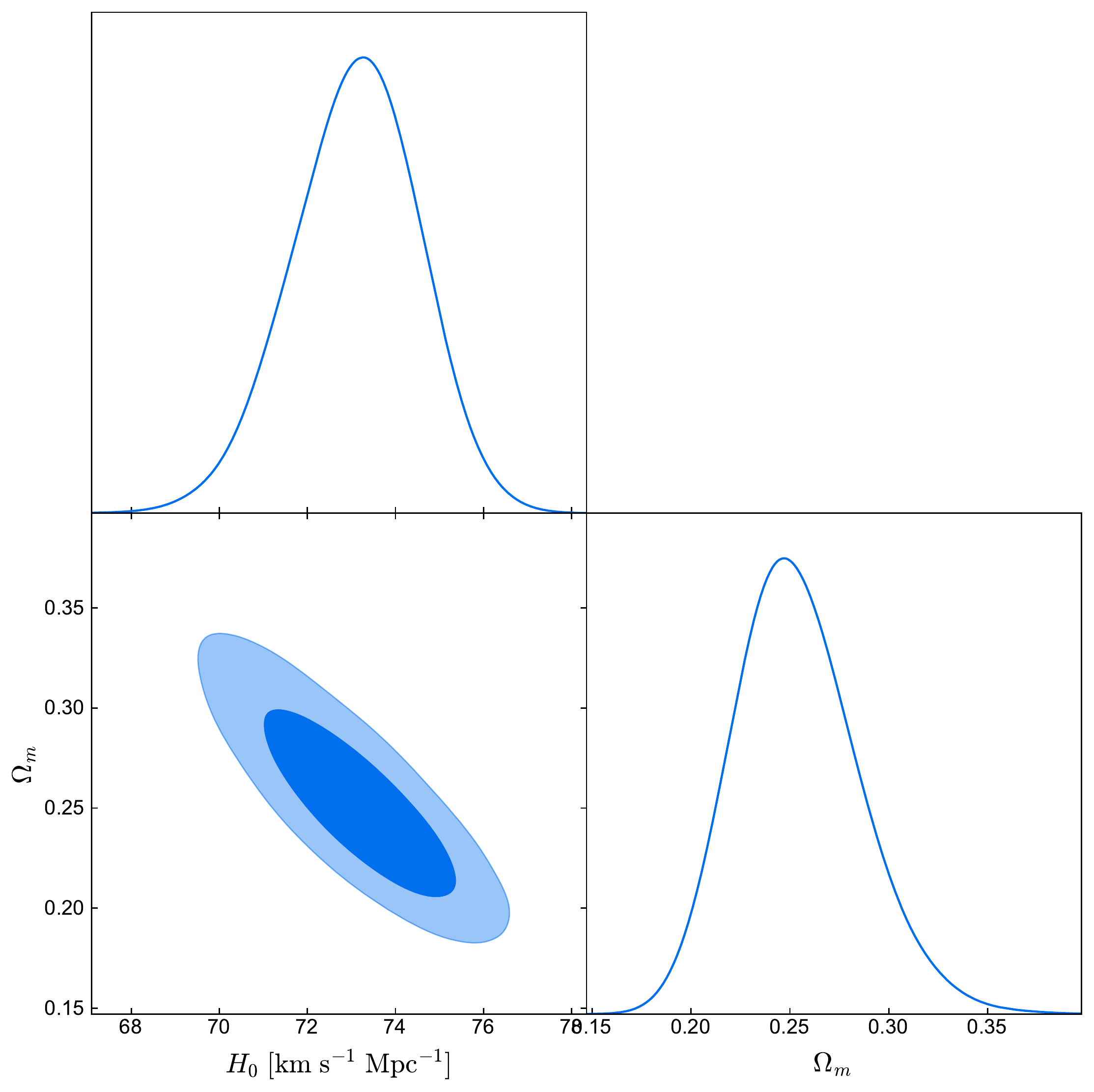}
\caption{1D and 2D marginalized probability distributions for the
parameters $H_0$ and $\Omega_K$ with strong lensing systems and
radio quasars, in the framework of flat $\Lambda$CDM
model.\label{LCDM}}
\end{figure}

\section{Results and discussion}

Implementing Python MCMC module EMCEE \cite{Foreman_Mackey_2013},
$\Omega_K$ and $H_0$ parameters are determined by maximizing the
final log-likelihood of
\begin{equation}
\mathrm{ln} {\cal L}= \mathrm{ln} ({\cal L}_{\rm{QSO}})+\mathrm{ln}
({\cal L}_{\Delta t})
\end{equation}
with the third-order polynomial coefficients ($a_1$, $a_2$) fitted
to the strong lensing and quasar data simultaneously. Performing
fits on the two parameters of interest ($\Omega_K$, $H_0$) and two
nuisance parameters ($a_1$, $a_2$) to the lensed + unlensed quasar
data, we obtain the results displayed in Table \ref{tabresult}. The
marginalized probability distribution of each parameter and the
marginalized 2D confidence contours are displayed in Fig.
\ref{Fig1}. The best-fit values with 68.3\% confidence level for the
four parameters are $H_0=78.3\pm2.9 \Mpc$, $\Omega_K=0.49\pm0.24$,
$a_1=-0.364\pm0.037$, and $a_2=0.068\pm0.014$. Compared with the
previous results obtained in a low redshift range
\cite{Collett:2019hrr}, our analysis results also demonstrate that
the cosmic curvature plays an important role in the determination of
the Hubble constant, which can be clearly seen from the positive
correlation between $\Omega_K$ and $H_0$ in Fig. \ref{Fig1}. Under
assumption of zero spatial curvature --- supported by other improved
model-independent methods referring to a distant past
\cite{cao2019milliarcsecond}, we get stringent constraints on the
Hubble constant as $H_0=73.6^{+1.8}_{-1.6} \Mpc$, with the
corresponding marginalized probability distribution presented in
Fig. \ref{Fig2}. In the framework of a non-flat and flat Universe,
the results for the lensed + unlensed quasar data are summarized in
Table \ref{tabresult}. Therefore, our model-independent $H_0$
constraints are well consistent with the recent determinations of
$H_0$ from the Supernovae $H_0$ for the SH0ES collaboration
\cite{Riess:2019cxk}, which is the most unambiguous result of the
current lensed + unlensed quasar dataset. Let us note that, at the
current observational level, quasars may achieve model-independent
$H_0$ measurements at much higher redshifts (which is especially
important in cosmology), compared with other popular astrophysical
probes (including SNe Ia) adopted as distance indicators for
providing the distance $d(z)$ \cite{Collett:2019hrr}.

Cosmic curvature $\Omega_K$ is a fundamental parameter for
cosmology. In this paper, we have focused on model-independent
measurement of $\Omega_K$ by applying distance sum rule to the
time-delay measurements of 7 strong lensing systems and 120
intermediate-luminosity quasars calibrated as standard rulers. With
the prior of $H_0=74.03 \Mpc$ determined locally via the distance
ladder, our final assessment of the cosmic curvature with
corresponding 1$\sigma$ uncertainty is $\Omega_K =
0.23^{+0.15}_{-0.17}$. The results are shown in Fig. \ref{Fig2},
which suggests that there is no significant signal indicating the
deviation of the cosmic curvature $\Omega_K$ from zero (spatially
flat geometry). Such unambiguous result of the available quasar
observations is also in agreement with the recent analysis focusing
on the source-lens/lens distance ratio in strong lensing systems
\cite{Qi:2018atg,Xia:2016dgk,Zhou:2019vou} and theoretical Hubble
diagram reconstructed by the Hubble parameter measurements
\cite{Liuyuting2020,Yu:2016gmd,Wang:2020dbt} in the framework of
other model-independent $\Omega_K$ test. The constraining power of
our method is more obvious when the large size difference between
the samples is taken into consideration.

\begin{table}[htb]
\centering
\begin{tabular}{cccc}
\hline
 & $H_0[\Mpc]$ &$\Omega_K$ & $\Omega_m$ \\
\hline
Non-flat $\Lambda$CDM & $72.5\pm1.6$ & $-0.09^{+0.13}_{-0.15}$ & $0.287\pm0.055$ \\
\hline
Flat $\Lambda$CDM & $73.1^{+1.5}_{-1.3}$ & $-$ & $0.254^{+0.027}_{-0.035}$  \\
\hline
\end{tabular}
\caption{Fitting results (68.3\% confidence level) for the open $\Lambda$CDM and flat $\Lambda$CDM.\label{tab2}}
\end{table}

Let us remark on two aspects. Firstly, from the observational point
of view, one can see that the 120 intermediate-luminosity quasars
have perfect coverage of source redshifts in 7 SGL systems ($z\sim
3$), acting as distance indicators for providing the distance $d(z)$
on the right side of Eq.~(\ref{timedelay1}). Therefore, we propose
to investigate the constraints on $H_0$ and $\Omega_K$ from the SGL
time delay data and radio quasars. Secondly, the constraint result
of $H_0$ from H0LiCOW is strongly dependent on cosmological models.
In order to investigate how sensitive our results on $H_0$ and
$\Omega_K$ are on the choice of cosmological model, we also perform
a comparative analysis of the current lensed + unlensed quasar
dataset in the non-flat and flat $\Lambda$CDM model. The results of
our parameter estimation computations are summarized in Table
\ref{tab2}, and with $1\sigma$ and $2\sigma$ confidence level
contours are shown in Fig. \ref{OLCDM}-\ref{LCDM}. From the combined
analyses with lensed + unlensed quasar dataset, we find that the
Hubble constant and the spatial curvature density parameter are
constrained to be $H_0=72.5\pm1.6 \Mpc$ and
$\Omega_K=-0.09^{+0.13}_{-0.15}$. In order to check the constraining
power of the lensed + unlensed quasar data on the Hubble constant,
we chose to assume zero spatial curvature and obtain
$H_0=73.1^{+1.5}_{-1.3} \Mpc$ in a flat universe. Now it is
worthwhile to make some comments on the results obtained above.
Firstly, it is interesting -- and might even be significant -- that
the $H_0$ and $\Omega_K$ constraints listed in Table \ref{tab2} are
quite consistent with estimates for these parameters from most other
data. More specifically, in broad terms, the estimated values of
$H_0$ are in agreement with the standard ones reported by the SH0ES
collaboration \cite{Riess:2019cxk}. Secondly, let us note that for
the flat and non-flat $\Lambda$CDM models, the matter density
parameter are fitted to $\Omega_m=0.287\pm0.055$ and
$\Omega_m=0.254^{+0.027}_{-0.035}$, respectively. Compared with
Planck fitting results \cite{Aghanim:2018eyx}, the best-fit values
for the present density parameters will considerably be improved,
with the help of the quasar observations.

In this paper, we focused on the idea of constraining $H_0$ and
$\Omega_K$ by using the observations of quasars: ultra-compact
structure in intermediate-luminosity radio quasars from the
very-long-baseline interferometry (VLBI) observations
\cite{Cao:2017ivt,Cao:2017abj,Zheng:2017asg,Cao:2016dgw,Qi:2017xzl,Liu:2020bac}
and the time-delay measurements of strong lensing systems with
quasars acting as background sources \cite{Wong:2019kwg}. Providing
a better redshift coverage of SGL systems, our method opens a new
possibility to quantitatively analyze current tensions concerning
the values of the Hubble constant and the curvature parameter with
multiple measurements of lensed and unlensed high-redshift quasars.
Note that there are many potential ways, in which our technique
might be improved by the discoverable lens population in future
surveys such as the Dark Energy Survey (DES), Vera C. Rubin
Observatory (Legacy Survey of Space and Time -- LSST), and Euclid.
For instance, LSST would enable the discovery of 3000 lensed quasars
in the most optimistic discovery scenario, with precise measurements
of time delays between multiple images
\cite{oguri2010gravitationally}. Our method could also be extended
to the SNe Ia-galaxy strong-lensing systems with exceptionally well
characterized spectral sequences and thus very accurate time delays
measured in lensed SNe Ia \cite{pereira2013}. Following the recent
analysis of the likely yields of LSST \cite{goldstein2016how},
there are 650 multiply imaged SNe Ia that could provide precise time
delays with supplementary data points on their light curves. On the
other hand, benefit from more recent VLBI imaging observations based
on better UV-coverage \cite{PK15}, both current and future VLBI
surveys will discover a large amount of intermediate-luminosity
radio quasars, with the angular sizes of the compact structure
observed at different frequencies \cite{cao2018cosmological}. In
summary, the approach introduced in this paper offers a new
model-independent way of simultaneously constraining both $H_0$ and
$\Omega_K$ at much higher accuracy, with future surveys of strongly
lensed quasars and high-quality radio astronomical observations of
quasars \cite{cao2020precise}.

\begin{acknowledgments}
This paper is dedicated to the 60th anniversary of the Department of
Astronomy, Beijing Normal University. This work was supported by the
Fundamental Research Funds for the Central Universities (Grant No.
N180503014); National Key R\&D Program of China No. 2017YFA0402600;
the National Natural Science Foundation of China under Grants Nos.
2021003, 11690023, and 11633001; Beijing Talents Fund of
Organization Department of Beijing Municipal Committee of the CPC;
the Strategic Priority Research Program of the Chinese Academy of
Sciences, Grant No. XDB23000000; the Interdiscipline Research Funds
of Beijing Normal University; and the Opening Project of Key
Laboratory of Computational Astrophysics, National Astronomical
Observatories, Chinese Academy of Sciences. M.B. was supported by
the Foreign Talent Introducing Project and Special Fund Support of
Foreign Knowledge Introducing Project in China. He was supported by
the Key Foreign Expert Program for the Central Universities No.
X2018002.
\end{acknowledgments}

\bibliography{HK_ref}

\begin{thebibliography}{62}
\expandafter\ifx\csname natexlab\endcsname\relax\def\natexlab#1{#1}\fi
\expandafter\ifx\csname bibnamefont\endcsname\relax
  \def\bibnamefont#1{#1}\fi
\expandafter\ifx\csname bibfnamefont\endcsname\relax
  \def\bibfnamefont#1{#1}\fi
\expandafter\ifx\csname citenamefont\endcsname\relax
  \def\citenamefont#1{#1}\fi
\expandafter\ifx\csname url\endcsname\relax
  \def\url#1{\texttt{#1}}\fi
\expandafter\ifx\csname urlprefix\endcsname\relax\def\urlprefix{URL }\fi
\providecommand{\bibinfo}[2]{#2}
\providecommand{\eprint}[2][]{\url{#2}}

\bibitem[{\citenamefont{Cao et~al.}(2010)\citenamefont{Cao, Liang, and
  Zhu}}]{cao2010testing}
\bibinfo{author}{\bibfnamefont{S.}~\bibnamefont{Cao}},
  \bibinfo{author}{\bibfnamefont{N.}~\bibnamefont{Liang}}, \bibnamefont{and}
  \bibinfo{author}{\bibfnamefont{Z.}~\bibnamefont{Zhu}},
  \bibinfo{journal}{Monthly Notices of the Royal Astronomical Society}
  \textbf{\bibinfo{volume}{416}}, \bibinfo{pages}{1099} (\bibinfo{year}{2010}).

\bibitem[{\citenamefont{Cao et~al.}(2012{\natexlab{a}})\citenamefont{Cao, Pan,
  Biesiada, Godlowski, and Zhu}}]{cao2012SL}
\bibinfo{author}{\bibfnamefont{S.}~\bibnamefont{Cao}},
  \bibinfo{author}{\bibfnamefont{Y.}~\bibnamefont{Pan}},
  \bibinfo{author}{\bibfnamefont{M.}~\bibnamefont{Biesiada}},
  \bibinfo{author}{\bibfnamefont{W.}~\bibnamefont{Godlowski}},
  \bibnamefont{and} \bibinfo{author}{\bibfnamefont{Z.}~\bibnamefont{Zhu}},
  \bibinfo{journal}{Journal of Cosmology and Astroparticle Physics}
  \textbf{\bibinfo{volume}{3}}, \bibinfo{pages}{16}
  (\bibinfo{year}{2012}{\natexlab{a}}).

\bibitem[{\citenamefont{Cao and Zhu}(2014)}]{cao2014cosmic}
\bibinfo{author}{\bibfnamefont{S.}~\bibnamefont{Cao}} \bibnamefont{and}
  \bibinfo{author}{\bibfnamefont{Z.-H.} \bibnamefont{Zhu}},
  \bibinfo{journal}{Physical Review D} \textbf{\bibinfo{volume}{90}},
  \bibinfo{pages}{083006} (\bibinfo{year}{2014}).

\bibitem[{\citenamefont{Cao et~al.}(2015)\citenamefont{Cao, Biesiada, Gavazzi,
  Pi{\'o}rkowska, and Zhu}}]{Cao:2015qja}
\bibinfo{author}{\bibfnamefont{S.}~\bibnamefont{Cao}},
  \bibinfo{author}{\bibfnamefont{M.}~\bibnamefont{Biesiada}},
  \bibinfo{author}{\bibfnamefont{R.}~\bibnamefont{Gavazzi}},
  \bibinfo{author}{\bibfnamefont{A.}~\bibnamefont{Pi{\'o}rkowska}},
  \bibnamefont{and} \bibinfo{author}{\bibfnamefont{Z.-H.} \bibnamefont{Zhu}},
  \bibinfo{journal}{The Astrophysical Journal} \textbf{\bibinfo{volume}{806}},
  \bibinfo{pages}{185} (\bibinfo{year}{2015}).

\bibitem[{\citenamefont{Aghanim et~al.}(2020)\citenamefont{Aghanim, Akrami,
  Ashdown, Aumont, Baccigalupi, Ballardini, Banday, Barreiro, Bartolo, Basak
  et~al.}}]{Aghanim:2018eyx}
\bibinfo{author}{\bibfnamefont{N.}~\bibnamefont{Aghanim}},
  \bibinfo{author}{\bibfnamefont{Y.}~\bibnamefont{Akrami}},
  \bibinfo{author}{\bibfnamefont{M.}~\bibnamefont{Ashdown}},
  \bibinfo{author}{\bibfnamefont{J.}~\bibnamefont{Aumont}},
  \bibinfo{author}{\bibfnamefont{C.}~\bibnamefont{Baccigalupi}},
  \bibinfo{author}{\bibfnamefont{M.}~\bibnamefont{Ballardini}},
  \bibinfo{author}{\bibfnamefont{A.}~\bibnamefont{Banday}},
  \bibinfo{author}{\bibfnamefont{R.}~\bibnamefont{Barreiro}},
  \bibinfo{author}{\bibfnamefont{N.}~\bibnamefont{Bartolo}},
  \bibinfo{author}{\bibfnamefont{S.}~\bibnamefont{Basak}},
  \bibnamefont{et~al.}, \bibinfo{journal}{Astronomy \& Astrophysics}
  \textbf{\bibinfo{volume}{641}}, \bibinfo{pages}{A6} (\bibinfo{year}{2020}).

\bibitem[{\citenamefont{Riess et~al.}(2019)\citenamefont{Riess, Casertano,
  Yuan, Macri, and Scolnic}}]{Riess:2019cxk}
\bibinfo{author}{\bibfnamefont{A.~G.} \bibnamefont{Riess}},
  \bibinfo{author}{\bibfnamefont{S.}~\bibnamefont{Casertano}},
  \bibinfo{author}{\bibfnamefont{W.}~\bibnamefont{Yuan}},
  \bibinfo{author}{\bibfnamefont{L.~M.} \bibnamefont{Macri}}, \bibnamefont{and}
  \bibinfo{author}{\bibfnamefont{D.}~\bibnamefont{Scolnic}},
  \bibinfo{journal}{Astrophys. J.} \textbf{\bibinfo{volume}{876}},
  \bibinfo{pages}{85} (\bibinfo{year}{2019}), \eprint{1903.07603}.

\bibitem[{\citenamefont{Di~Valentino et~al.}(2018)\citenamefont{Di~Valentino,
  Melchiorri, Fantaye, and Heavens}}]{DiValentino:2018zjj}
\bibinfo{author}{\bibfnamefont{E.}~\bibnamefont{Di~Valentino}},
  \bibinfo{author}{\bibfnamefont{A.}~\bibnamefont{Melchiorri}},
  \bibinfo{author}{\bibfnamefont{Y.}~\bibnamefont{Fantaye}}, \bibnamefont{and}
  \bibinfo{author}{\bibfnamefont{A.}~\bibnamefont{Heavens}},
  \bibinfo{journal}{Phys. Rev. D} \textbf{\bibinfo{volume}{98}},
  \bibinfo{pages}{063508} (\bibinfo{year}{2018}), \eprint{1808.09201}.

\bibitem[{\citenamefont{Follin and Knox}(2018)}]{Follin:2017ljs}
\bibinfo{author}{\bibfnamefont{B.}~\bibnamefont{Follin}} \bibnamefont{and}
  \bibinfo{author}{\bibfnamefont{L.}~\bibnamefont{Knox}},
  \bibinfo{journal}{Mon. Not. Roy. Astron. Soc.}
  \textbf{\bibinfo{volume}{477}}, \bibinfo{pages}{4534} (\bibinfo{year}{2018}),
  \eprint{1707.01175}.

\bibitem[{\citenamefont{Refsdal}(1964)}]{Refsdal:1964nw}
\bibinfo{author}{\bibfnamefont{S.}~\bibnamefont{Refsdal}},
  \bibinfo{journal}{Mon. Not. Roy. Astron. Soc.}
  \textbf{\bibinfo{volume}{128}}, \bibinfo{pages}{307} (\bibinfo{year}{1964}).

\bibitem[{\citenamefont{Kelly et~al.}(2015)}]{Kelly:2014mwa}
\bibinfo{author}{\bibfnamefont{P.~L.} \bibnamefont{Kelly}}
  \bibnamefont{et~al.}, \bibinfo{journal}{Science}
  \textbf{\bibinfo{volume}{347}}, \bibinfo{pages}{1123} (\bibinfo{year}{2015}),
  \eprint{1411.6009}.

\bibitem[{\citenamefont{Goobar et~al.}(2017)}]{Goobar:2016uuf}
\bibinfo{author}{\bibfnamefont{A.}~\bibnamefont{Goobar}} \bibnamefont{et~al.},
  \bibinfo{journal}{Science} \textbf{\bibinfo{volume}{356}},
  \bibinfo{pages}{291} (\bibinfo{year}{2017}), \eprint{1611.00014}.

\bibitem[{\citenamefont{Wong et~al.}(2020)\citenamefont{Wong, Suyu, Chen, Rusu,
  Millon, Sluse, Bonvin, Fassnacht, Taubenberger, Auger et~al.}}]{Wong:2019kwg}
\bibinfo{author}{\bibfnamefont{K.~C.} \bibnamefont{Wong}},
  \bibinfo{author}{\bibfnamefont{S.~H.} \bibnamefont{Suyu}},
  \bibinfo{author}{\bibfnamefont{G.~C.} \bibnamefont{Chen}},
  \bibinfo{author}{\bibfnamefont{C.~E.} \bibnamefont{Rusu}},
  \bibinfo{author}{\bibfnamefont{M.}~\bibnamefont{Millon}},
  \bibinfo{author}{\bibfnamefont{D.}~\bibnamefont{Sluse}},
  \bibinfo{author}{\bibfnamefont{V.}~\bibnamefont{Bonvin}},
  \bibinfo{author}{\bibfnamefont{C.~D.} \bibnamefont{Fassnacht}},
  \bibinfo{author}{\bibfnamefont{S.}~\bibnamefont{Taubenberger}},
  \bibinfo{author}{\bibfnamefont{M.~W.} \bibnamefont{Auger}},
  \bibnamefont{et~al.}, \bibinfo{journal}{Monthly Notices of the Royal
  Astronomical Society} \textbf{\bibinfo{volume}{498}}, \bibinfo{pages}{1420}
  (\bibinfo{year}{2020}).

\bibitem[{\citenamefont{Di~Valentino et~al.}(2019)\citenamefont{Di~Valentino,
  Melchiorri, and Silk}}]{DiValentino:2019qzk}
\bibinfo{author}{\bibfnamefont{E.}~\bibnamefont{Di~Valentino}},
  \bibinfo{author}{\bibfnamefont{A.}~\bibnamefont{Melchiorri}},
  \bibnamefont{and} \bibinfo{author}{\bibfnamefont{J.}~\bibnamefont{Silk}},
  \bibinfo{journal}{Nature Astron.} \textbf{\bibinfo{volume}{4}},
  \bibinfo{pages}{196} (\bibinfo{year}{2019}), \eprint{1911.02087}.

\bibitem[{\citenamefont{Di~Valentino et~al.}(2020)\citenamefont{Di~Valentino,
  Melchiorri, and Silk}}]{DiValentino:2020hov}
\bibinfo{author}{\bibfnamefont{E.}~\bibnamefont{Di~Valentino}},
  \bibinfo{author}{\bibfnamefont{A.}~\bibnamefont{Melchiorri}},
  \bibnamefont{and} \bibinfo{author}{\bibfnamefont{J.}~\bibnamefont{Silk}},
  \bibinfo{journal}{arXiv preprint arXiv:2003.04935}  (\bibinfo{year}{2020}).

\bibitem[{\citenamefont{{Handley}}(2019)}]{2019arXiv190809139H}
\bibinfo{author}{\bibfnamefont{W.}~\bibnamefont{{Handley}}},
  \bibinfo{journal}{arXiv e-prints} \bibinfo{eid}{arXiv:1908.09139}
  (\bibinfo{year}{2019}), \eprint{1908.09139}.

\bibitem[{\citenamefont{Guo et~al.}(2019)\citenamefont{Guo, Zhang, and
  Zhang}}]{guo2019can}
\bibinfo{author}{\bibfnamefont{R.-Y.} \bibnamefont{Guo}},
  \bibinfo{author}{\bibfnamefont{J.-F.} \bibnamefont{Zhang}}, \bibnamefont{and}
  \bibinfo{author}{\bibfnamefont{X.}~\bibnamefont{Zhang}},
  \bibinfo{journal}{Journal of Cosmology and Astroparticle Physics}
  \textbf{\bibinfo{volume}{2019}}, \bibinfo{pages}{054} (\bibinfo{year}{2019}).

\bibitem[{\citenamefont{Vagnozzi}(2020)}]{vagnozzi2020new}
\bibinfo{author}{\bibfnamefont{S.}~\bibnamefont{Vagnozzi}},
  \bibinfo{journal}{Physical Review D} \textbf{\bibinfo{volume}{102}},
  \bibinfo{pages}{023518} (\bibinfo{year}{2020}).

\bibitem[{\citenamefont{Zhang}(2019)}]{zhang2019gravitational}
\bibinfo{author}{\bibfnamefont{X.}~\bibnamefont{Zhang}},
  \bibinfo{journal}{SCIENCE CHINA Physics, Mechanics \& Astronomy}
  \textbf{\bibinfo{volume}{62}}, \bibinfo{pages}{110431}
  (\bibinfo{year}{2019}).

\bibitem[{\citenamefont{Qi and Zhang}(2020)}]{qi2020new}
\bibinfo{author}{\bibfnamefont{J.-Z.} \bibnamefont{Qi}} \bibnamefont{and}
  \bibinfo{author}{\bibfnamefont{X.}~\bibnamefont{Zhang}},
  \bibinfo{journal}{Chinese Physics C} \textbf{\bibinfo{volume}{44}},
  \bibinfo{pages}{055101} (\bibinfo{year}{2020}).

\bibitem[{\citenamefont{Vattis et~al.}(2019)\citenamefont{Vattis, Koushiappas,
  and Loeb}}]{vattis2019dark}
\bibinfo{author}{\bibfnamefont{K.}~\bibnamefont{Vattis}},
  \bibinfo{author}{\bibfnamefont{S.~M.} \bibnamefont{Koushiappas}},
  \bibnamefont{and} \bibinfo{author}{\bibfnamefont{A.}~\bibnamefont{Loeb}},
  \bibinfo{journal}{Physical Review D} \textbf{\bibinfo{volume}{99}},
  \bibinfo{pages}{121302} (\bibinfo{year}{2019}).

\bibitem[{\citenamefont{Zhang et~al.}(2014)\citenamefont{Zhang, Geng, and
  Zhang}}]{zhang2014neutrinos}
\bibinfo{author}{\bibfnamefont{J.-F.} \bibnamefont{Zhang}},
  \bibinfo{author}{\bibfnamefont{J.-J.} \bibnamefont{Geng}}, \bibnamefont{and}
  \bibinfo{author}{\bibfnamefont{X.}~\bibnamefont{Zhang}},
  \bibinfo{journal}{Journal of Cosmology and Astroparticle Physics}
  \textbf{\bibinfo{volume}{2014}}, \bibinfo{pages}{044} (\bibinfo{year}{2014}).

\bibitem[{\citenamefont{Zhao et~al.}(2017)\citenamefont{Zhao, He, Zhang, and
  Zhang}}]{zhao2017search}
\bibinfo{author}{\bibfnamefont{M.-M.} \bibnamefont{Zhao}},
  \bibinfo{author}{\bibfnamefont{D.-Z.} \bibnamefont{He}},
  \bibinfo{author}{\bibfnamefont{J.-F.} \bibnamefont{Zhang}}, \bibnamefont{and}
  \bibinfo{author}{\bibfnamefont{X.}~\bibnamefont{Zhang}},
  \bibinfo{journal}{Physical Review D} \textbf{\bibinfo{volume}{96}},
  \bibinfo{pages}{043520} (\bibinfo{year}{2017}).

\bibitem[{\citenamefont{Guo and Zhang}(2017)}]{guo2017constraints}
\bibinfo{author}{\bibfnamefont{R.-Y.} \bibnamefont{Guo}} \bibnamefont{and}
  \bibinfo{author}{\bibfnamefont{X.}~\bibnamefont{Zhang}},
  \bibinfo{journal}{The European Physical Journal C}
  \textbf{\bibinfo{volume}{77}}, \bibinfo{pages}{882} (\bibinfo{year}{2017}).

\bibitem[{\citenamefont{Rasanen et~al.}(2015)\citenamefont{Rasanen, Bolejko,
  and Finoguenov}}]{Rasanen:2014mca}
\bibinfo{author}{\bibfnamefont{S.}~\bibnamefont{Rasanen}},
  \bibinfo{author}{\bibfnamefont{K.}~\bibnamefont{Bolejko}}, \bibnamefont{and}
  \bibinfo{author}{\bibfnamefont{A.}~\bibnamefont{Finoguenov}},
  \bibinfo{journal}{Phys. Rev. Lett.} \textbf{\bibinfo{volume}{115}},
  \bibinfo{pages}{101301} (\bibinfo{year}{2015}), \eprint{1412.4976}.

\bibitem[{\citenamefont{Qi et~al.}(2019{\natexlab{a}})\citenamefont{Qi, Cao,
  Biesiada, Ding, Zhu, and Zheng}}]{Qi:2018atg}
\bibinfo{author}{\bibfnamefont{J.}~\bibnamefont{Qi}},
  \bibinfo{author}{\bibfnamefont{S.}~\bibnamefont{Cao}},
  \bibinfo{author}{\bibfnamefont{M.}~\bibnamefont{Biesiada}},
  \bibinfo{author}{\bibfnamefont{X.}~\bibnamefont{Ding}},
  \bibinfo{author}{\bibfnamefont{Z.-H.} \bibnamefont{Zhu}}, \bibnamefont{and}
  \bibinfo{author}{\bibfnamefont{X.}~\bibnamefont{Zheng}},
  \bibinfo{journal}{Phys. Rev. D} \textbf{\bibinfo{volume}{100}},
  \bibinfo{pages}{023530} (\bibinfo{year}{2019}{\natexlab{a}}),
  \eprint{1802.05532}.

\bibitem[{\citenamefont{Wang et~al.}(2020{\natexlab{a}})\citenamefont{Wang, Qi,
  Zhang, and Zhang}}]{Wang:2019yob}
\bibinfo{author}{\bibfnamefont{B.}~\bibnamefont{Wang}},
  \bibinfo{author}{\bibfnamefont{J.-Z.} \bibnamefont{Qi}},
  \bibinfo{author}{\bibfnamefont{J.-F.} \bibnamefont{Zhang}}, \bibnamefont{and}
  \bibinfo{author}{\bibfnamefont{X.}~\bibnamefont{Zhang}},
  \bibinfo{journal}{The Astrophysical Journal} \textbf{\bibinfo{volume}{898}},
  \bibinfo{pages}{100} (\bibinfo{year}{2020}{\natexlab{a}}).

\bibitem[{\citenamefont{Xia et~al.}(2017)\citenamefont{Xia, Yu, Wang, Tian, Li,
  Cao, and Zhu}}]{Xia:2016dgk}
\bibinfo{author}{\bibfnamefont{J.-Q.} \bibnamefont{Xia}},
  \bibinfo{author}{\bibfnamefont{H.}~\bibnamefont{Yu}},
  \bibinfo{author}{\bibfnamefont{G.-J.} \bibnamefont{Wang}},
  \bibinfo{author}{\bibfnamefont{S.-X.} \bibnamefont{Tian}},
  \bibinfo{author}{\bibfnamefont{Z.-X.} \bibnamefont{Li}},
  \bibinfo{author}{\bibfnamefont{S.}~\bibnamefont{Cao}}, \bibnamefont{and}
  \bibinfo{author}{\bibfnamefont{Z.-H.} \bibnamefont{Zhu}},
  \bibinfo{journal}{Astrophys. J.} \textbf{\bibinfo{volume}{834}},
  \bibinfo{pages}{75} (\bibinfo{year}{2017}), \eprint{1611.04731}.

\bibitem[{\citenamefont{Li et~al.}(2018)\citenamefont{Li, Ding, Wang, Liao, and
  Zhu}}]{Li:2018hyr}
\bibinfo{author}{\bibfnamefont{Z.}~\bibnamefont{Li}},
  \bibinfo{author}{\bibfnamefont{X.}~\bibnamefont{Ding}},
  \bibinfo{author}{\bibfnamefont{G.-J.} \bibnamefont{Wang}},
  \bibinfo{author}{\bibfnamefont{K.}~\bibnamefont{Liao}}, \bibnamefont{and}
  \bibinfo{author}{\bibfnamefont{Z.-H.} \bibnamefont{Zhu}},
  \bibinfo{journal}{Astrophys. J.} \textbf{\bibinfo{volume}{854}},
  \bibinfo{pages}{146} (\bibinfo{year}{2018}), \eprint{1801.08001}.

\bibitem[{\citenamefont{Zhou and Li}(2020)}]{Zhou:2019vou}
\bibinfo{author}{\bibfnamefont{H.}~\bibnamefont{Zhou}} \bibnamefont{and}
  \bibinfo{author}{\bibfnamefont{Z.-X.} \bibnamefont{Li}},
  \bibinfo{journal}{Astrophys. J.} \textbf{\bibinfo{volume}{899}},
  \bibinfo{pages}{186} (\bibinfo{year}{2020}), \eprint{1912.01828}.

\bibitem[{\citenamefont{Chen et~al.}(2019)\citenamefont{Chen, Li, Shu, and
  Cao}}]{Chen:2018jcf}
\bibinfo{author}{\bibfnamefont{Y.}~\bibnamefont{Chen}},
  \bibinfo{author}{\bibfnamefont{R.}~\bibnamefont{Li}},
  \bibinfo{author}{\bibfnamefont{Y.}~\bibnamefont{Shu}}, \bibnamefont{and}
  \bibinfo{author}{\bibfnamefont{X.}~\bibnamefont{Cao}}, \bibinfo{journal}{Mon.
  Not. Roy. Astron. Soc.} \textbf{\bibinfo{volume}{488}}, \bibinfo{pages}{3745}
  (\bibinfo{year}{2019}), \eprint{1809.09845}.

\bibitem[{\citenamefont{Qi et~al.}(2019{\natexlab{b}})\citenamefont{Qi, Cao,
  Zhang, Biesiada, Wu, and Zhu}}]{Qi:2018aio}
\bibinfo{author}{\bibfnamefont{J.-Z.} \bibnamefont{Qi}},
  \bibinfo{author}{\bibfnamefont{S.}~\bibnamefont{Cao}},
  \bibinfo{author}{\bibfnamefont{S.}~\bibnamefont{Zhang}},
  \bibinfo{author}{\bibfnamefont{M.}~\bibnamefont{Biesiada}},
  \bibinfo{author}{\bibfnamefont{Y.}~\bibnamefont{Wu}}, \bibnamefont{and}
  \bibinfo{author}{\bibfnamefont{Z.-H.} \bibnamefont{Zhu}},
  \bibinfo{journal}{Mon. Not. Roy. Astron. Soc.}
  \textbf{\bibinfo{volume}{483}}, \bibinfo{pages}{1104}
  (\bibinfo{year}{2019}{\natexlab{b}}), \eprint{1803.01990}.

\bibitem[{\citenamefont{Liao}(2019)}]{Liao:2019hfl}
\bibinfo{author}{\bibfnamefont{K.}~\bibnamefont{Liao}}, \bibinfo{journal}{Phys.
  Rev. D} \textbf{\bibinfo{volume}{99}}, \bibinfo{pages}{083514}
  (\bibinfo{year}{2019}), \eprint{1904.01744}.

\bibitem[{\citenamefont{Cao et~al.}(2012{\natexlab{b}})\citenamefont{Cao, Pan,
  Biesiada, Godlowski, and Zhu}}]{cao2012constraints}
\bibinfo{author}{\bibfnamefont{S.}~\bibnamefont{Cao}},
  \bibinfo{author}{\bibfnamefont{Y.}~\bibnamefont{Pan}},
  \bibinfo{author}{\bibfnamefont{M.}~\bibnamefont{Biesiada}},
  \bibinfo{author}{\bibfnamefont{W.}~\bibnamefont{Godlowski}},
  \bibnamefont{and} \bibinfo{author}{\bibfnamefont{Z.}~\bibnamefont{Zhu}},
  \bibinfo{journal}{Journal of Cosmology and Astroparticle Physics}
  \textbf{\bibinfo{volume}{2012}}, \bibinfo{pages}{016}
  (\bibinfo{year}{2012}{\natexlab{b}}).

\bibitem[{\citenamefont{Collett et~al.}(2019)\citenamefont{Collett, Montanari,
  and Rasanen}}]{Collett:2019hrr}
\bibinfo{author}{\bibfnamefont{T.}~\bibnamefont{Collett}},
  \bibinfo{author}{\bibfnamefont{F.}~\bibnamefont{Montanari}},
  \bibnamefont{and} \bibinfo{author}{\bibfnamefont{S.}~\bibnamefont{Rasanen}},
  \bibinfo{journal}{Phys. Rev. Lett.} \textbf{\bibinfo{volume}{123}},
  \bibinfo{pages}{231101} (\bibinfo{year}{2019}), \eprint{1905.09781}.

\bibitem[{\citenamefont{Wei and Melia}(2020)}]{Wei:2020suh}
\bibinfo{author}{\bibfnamefont{J.-J.} \bibnamefont{Wei}} \bibnamefont{and}
  \bibinfo{author}{\bibfnamefont{F.}~\bibnamefont{Melia}},
  \bibinfo{journal}{arXiv preprint arXiv:2005.10422}  (\bibinfo{year}{2020}).

\bibitem[{\citenamefont{Cao et~al.}(2019{\natexlab{a}})\citenamefont{Cao, Qi,
  Cao, Biesiada, Li, Pan, and Zhu}}]{cao2019direct}
\bibinfo{author}{\bibfnamefont{S.}~\bibnamefont{Cao}},
  \bibinfo{author}{\bibfnamefont{J.}~\bibnamefont{Qi}},
  \bibinfo{author}{\bibfnamefont{Z.}~\bibnamefont{Cao}},
  \bibinfo{author}{\bibfnamefont{M.}~\bibnamefont{Biesiada}},
  \bibinfo{author}{\bibfnamefont{J.}~\bibnamefont{Li}},
  \bibinfo{author}{\bibfnamefont{Y.}~\bibnamefont{Pan}}, \bibnamefont{and}
  \bibinfo{author}{\bibfnamefont{Z.}~\bibnamefont{Zhu}},
  \bibinfo{journal}{Scientific Reports} \textbf{\bibinfo{volume}{9}},
  \bibinfo{pages}{11608} (\bibinfo{year}{2019}{\natexlab{a}}).

\bibitem[{\citenamefont{Shajib et~al.}(2020)}]{shajib2019strides}
\bibinfo{author}{\bibfnamefont{A.}~\bibnamefont{Shajib}} \bibnamefont{et~al.}
  (\bibinfo{collaboration}{DES}), \bibinfo{journal}{Mon. Not. Roy. Astron.
  Soc.} \textbf{\bibinfo{volume}{494}}, \bibinfo{pages}{6072}
  (\bibinfo{year}{2020}), \eprint{1910.06306}.

\bibitem[{\citenamefont{{Suyu} et~al.}(2010)\citenamefont{{Suyu}, {Marshall},
  {Auger}, {Hilbert}, {Blandford}, {Koopmans}, {Fassnacht}, and
  {Treu}}}]{Suyu10}
\bibinfo{author}{\bibfnamefont{S.~H.} \bibnamefont{{Suyu}}},
  \bibinfo{author}{\bibfnamefont{P.~J.} \bibnamefont{{Marshall}}},
  \bibinfo{author}{\bibfnamefont{M.~W.} \bibnamefont{{Auger}}},
  \bibinfo{author}{\bibfnamefont{S.}~\bibnamefont{{Hilbert}}},
  \bibinfo{author}{\bibfnamefont{R.~D.} \bibnamefont{{Blandford}}},
  \bibinfo{author}{\bibfnamefont{L.~V.~E.} \bibnamefont{{Koopmans}}},
  \bibinfo{author}{\bibfnamefont{C.~D.} \bibnamefont{{Fassnacht}}},
  \bibnamefont{and} \bibinfo{author}{\bibfnamefont{T.}~\bibnamefont{{Treu}}},
  \bibinfo{journal}{The Astrophysical Journal} \textbf{\bibinfo{volume}{711}},
  \bibinfo{pages}{201} (\bibinfo{year}{2010}), \eprint{0910.2773}.

\bibitem[{\citenamefont{{Jee} et~al.}(2019)\citenamefont{{Jee}, {Suyu},
  {Komatsu}, {Fassnacht}, {Hilbert}, and {Koopmans}}}]{Jee19}
\bibinfo{author}{\bibfnamefont{I.}~\bibnamefont{{Jee}}},
  \bibinfo{author}{\bibfnamefont{S.~H.} \bibnamefont{{Suyu}}},
  \bibinfo{author}{\bibfnamefont{E.}~\bibnamefont{{Komatsu}}},
  \bibinfo{author}{\bibfnamefont{C.~D.} \bibnamefont{{Fassnacht}}},
  \bibinfo{author}{\bibfnamefont{S.}~\bibnamefont{{Hilbert}}},
  \bibnamefont{and} \bibinfo{author}{\bibfnamefont{L.~V.~E.}
  \bibnamefont{{Koopmans}}}, \bibinfo{journal}{Science}
  \textbf{\bibinfo{volume}{365}}, \bibinfo{pages}{1134} (\bibinfo{year}{2019}),
  \eprint{1909.06712}.

\bibitem[{\citenamefont{{Suyu} et~al.}(2013)\citenamefont{{Suyu}, {Auger},
  {Hilbert}, {Marshall}, {Tewes}, {Treu}, {Fassnacht}, {Koopmans}, {Sluse},
  {Bland ford} et~al.}}]{Suyu13}
\bibinfo{author}{\bibfnamefont{S.~H.} \bibnamefont{{Suyu}}},
  \bibinfo{author}{\bibfnamefont{M.~W.} \bibnamefont{{Auger}}},
  \bibinfo{author}{\bibfnamefont{S.}~\bibnamefont{{Hilbert}}},
  \bibinfo{author}{\bibfnamefont{P.~J.} \bibnamefont{{Marshall}}},
  \bibinfo{author}{\bibfnamefont{M.}~\bibnamefont{{Tewes}}},
  \bibinfo{author}{\bibfnamefont{T.}~\bibnamefont{{Treu}}},
  \bibinfo{author}{\bibfnamefont{C.~D.} \bibnamefont{{Fassnacht}}},
  \bibinfo{author}{\bibfnamefont{L.~V.~E.} \bibnamefont{{Koopmans}}},
  \bibinfo{author}{\bibfnamefont{D.}~\bibnamefont{{Sluse}}},
  \bibinfo{author}{\bibfnamefont{R.~D.} \bibnamefont{{Bland ford}}},
  \bibnamefont{et~al.}, \bibinfo{journal}{The Astrophysical Journal}
  \textbf{\bibinfo{volume}{766}}, \bibinfo{eid}{70} (\bibinfo{year}{2013}),
  \eprint{1208.6010}.

\bibitem[{\citenamefont{{Suyu} et~al.}(2014)\citenamefont{{Suyu}, {Treu},
  {Hilbert}, {Sonnenfeld}, {Auger}, {Blandford}, {Collett}, {Courbin},
  {Fassnacht}, {Koopmans} et~al.}}]{Suyu14}
\bibinfo{author}{\bibfnamefont{S.~H.} \bibnamefont{{Suyu}}},
  \bibinfo{author}{\bibfnamefont{T.}~\bibnamefont{{Treu}}},
  \bibinfo{author}{\bibfnamefont{S.}~\bibnamefont{{Hilbert}}},
  \bibinfo{author}{\bibfnamefont{A.}~\bibnamefont{{Sonnenfeld}}},
  \bibinfo{author}{\bibfnamefont{M.~W.} \bibnamefont{{Auger}}},
  \bibinfo{author}{\bibfnamefont{R.~D.} \bibnamefont{{Blandford}}},
  \bibinfo{author}{\bibfnamefont{T.}~\bibnamefont{{Collett}}},
  \bibinfo{author}{\bibfnamefont{F.}~\bibnamefont{{Courbin}}},
  \bibinfo{author}{\bibfnamefont{C.~D.} \bibnamefont{{Fassnacht}}},
  \bibinfo{author}{\bibfnamefont{L.~V.~E.} \bibnamefont{{Koopmans}}},
  \bibnamefont{et~al.}, \bibinfo{journal}{The Astrophysical Journal Letters}
  \textbf{\bibinfo{volume}{788}}, \bibinfo{eid}{L35} (\bibinfo{year}{2014}),
  \eprint{1306.4732}.

\bibitem[{\citenamefont{{Chen} et~al.}(2019)\citenamefont{{Chen}, {Fassnacht},
  {Suyu}, {Rusu}, {Chan}, {Wong}, {Auger}, {Hilbert}, {Bonvin}, {Birrer}
  et~al.}}]{Chen19}
\bibinfo{author}{\bibfnamefont{G.~C.~F.} \bibnamefont{{Chen}}},
  \bibinfo{author}{\bibfnamefont{C.~D.} \bibnamefont{{Fassnacht}}},
  \bibinfo{author}{\bibfnamefont{S.~H.} \bibnamefont{{Suyu}}},
  \bibinfo{author}{\bibfnamefont{C.~E.} \bibnamefont{{Rusu}}},
  \bibinfo{author}{\bibfnamefont{J.~H.~H.} \bibnamefont{{Chan}}},
  \bibinfo{author}{\bibfnamefont{K.~C.} \bibnamefont{{Wong}}},
  \bibinfo{author}{\bibfnamefont{M.~W.} \bibnamefont{{Auger}}},
  \bibinfo{author}{\bibfnamefont{S.}~\bibnamefont{{Hilbert}}},
  \bibinfo{author}{\bibfnamefont{V.}~\bibnamefont{{Bonvin}}},
  \bibinfo{author}{\bibfnamefont{S.}~\bibnamefont{{Birrer}}},
  \bibnamefont{et~al.}, \bibinfo{journal}{Monthly Notices of the Royal
  Astronomical Society} \textbf{\bibinfo{volume}{490}}, \bibinfo{pages}{1743}
  (\bibinfo{year}{2019}), \eprint{1907.02533}.

\bibitem[{\citenamefont{{Wong} et~al.}(2017)\citenamefont{{Wong}, {Suyu},
  {Auger}, {Bonvin}, {Courbin}, {Fassnacht}, {Halkola}, {Rusu}, {Sluse},
  {Sonnenfeld} et~al.}}]{Wong17}
\bibinfo{author}{\bibfnamefont{K.~C.} \bibnamefont{{Wong}}},
  \bibinfo{author}{\bibfnamefont{S.~H.} \bibnamefont{{Suyu}}},
  \bibinfo{author}{\bibfnamefont{M.~W.} \bibnamefont{{Auger}}},
  \bibinfo{author}{\bibfnamefont{V.}~\bibnamefont{{Bonvin}}},
  \bibinfo{author}{\bibfnamefont{F.}~\bibnamefont{{Courbin}}},
  \bibinfo{author}{\bibfnamefont{C.~D.} \bibnamefont{{Fassnacht}}},
  \bibinfo{author}{\bibfnamefont{A.}~\bibnamefont{{Halkola}}},
  \bibinfo{author}{\bibfnamefont{C.~E.} \bibnamefont{{Rusu}}},
  \bibinfo{author}{\bibfnamefont{D.}~\bibnamefont{{Sluse}}},
  \bibinfo{author}{\bibfnamefont{A.~r.} \bibnamefont{{Sonnenfeld}}},
  \bibnamefont{et~al.}, \bibinfo{journal}{Monthly Notices of the Royal
  Astronomical Society} \textbf{\bibinfo{volume}{465}}, \bibinfo{pages}{4895}
  (\bibinfo{year}{2017}), \eprint{1607.01403}.

\bibitem[{\citenamefont{{Birrer} et~al.}(2019)\citenamefont{{Birrer}, {Treu},
  {Rusu}, {Bonvin}, {Fassnacht}, {Chan}, {Agnello}, {Shajib}, {Chen}, {Auger}
  et~al.}}]{Birrer19}
\bibinfo{author}{\bibfnamefont{S.}~\bibnamefont{{Birrer}}},
  \bibinfo{author}{\bibfnamefont{T.}~\bibnamefont{{Treu}}},
  \bibinfo{author}{\bibfnamefont{C.~E.} \bibnamefont{{Rusu}}},
  \bibinfo{author}{\bibfnamefont{V.}~\bibnamefont{{Bonvin}}},
  \bibinfo{author}{\bibfnamefont{C.~D.} \bibnamefont{{Fassnacht}}},
  \bibinfo{author}{\bibfnamefont{J.~H.~H.} \bibnamefont{{Chan}}},
  \bibinfo{author}{\bibfnamefont{A.}~\bibnamefont{{Agnello}}},
  \bibinfo{author}{\bibfnamefont{A.~J.} \bibnamefont{{Shajib}}},
  \bibinfo{author}{\bibfnamefont{G.~C.~F.} \bibnamefont{{Chen}}},
  \bibinfo{author}{\bibfnamefont{M.}~\bibnamefont{{Auger}}},
  \bibnamefont{et~al.}, \bibinfo{journal}{Monthly Notices of the Royal
  Astronomical Society} \textbf{\bibinfo{volume}{484}}, \bibinfo{pages}{4726}
  (\bibinfo{year}{2019}), \eprint{1809.01274}.

\bibitem[{\citenamefont{{Rusu} et~al.}(2020)\citenamefont{{Rusu}, {Wong},
  {Bonvin}, {Sluse}, {Suyu}, {Fassnacht}, {Chan}, {Hilbert}, {Auger},
  {Sonnenfeld} et~al.}}]{Rusu20}
\bibinfo{author}{\bibfnamefont{C.~E.} \bibnamefont{{Rusu}}},
  \bibinfo{author}{\bibfnamefont{K.~C.} \bibnamefont{{Wong}}},
  \bibinfo{author}{\bibfnamefont{V.}~\bibnamefont{{Bonvin}}},
  \bibinfo{author}{\bibfnamefont{D.}~\bibnamefont{{Sluse}}},
  \bibinfo{author}{\bibfnamefont{S.~H.} \bibnamefont{{Suyu}}},
  \bibinfo{author}{\bibfnamefont{C.~D.} \bibnamefont{{Fassnacht}}},
  \bibinfo{author}{\bibfnamefont{J.~H.~H.} \bibnamefont{{Chan}}},
  \bibinfo{author}{\bibfnamefont{S.}~\bibnamefont{{Hilbert}}},
  \bibinfo{author}{\bibfnamefont{M.~W.} \bibnamefont{{Auger}}},
  \bibinfo{author}{\bibfnamefont{A.}~\bibnamefont{{Sonnenfeld}}},
  \bibnamefont{et~al.}, \bibinfo{journal}{Monthly Notices of the Royal
  Astronomical Society}  (\bibinfo{year}{2020}), \eprint{1905.09338}.

\bibitem[{\citenamefont{Cao et~al.}(2017{\natexlab{a}})\citenamefont{Cao,
  Biesiada, Jackson, Zheng, Zhao, and Zhu}}]{Cao:2016dgw}
\bibinfo{author}{\bibfnamefont{S.}~\bibnamefont{Cao}},
  \bibinfo{author}{\bibfnamefont{M.}~\bibnamefont{Biesiada}},
  \bibinfo{author}{\bibfnamefont{J.}~\bibnamefont{Jackson}},
  \bibinfo{author}{\bibfnamefont{X.}~\bibnamefont{Zheng}},
  \bibinfo{author}{\bibfnamefont{Y.}~\bibnamefont{Zhao}}, \bibnamefont{and}
  \bibinfo{author}{\bibfnamefont{Z.-H.} \bibnamefont{Zhu}},
  \bibinfo{journal}{JCAP} \textbf{\bibinfo{volume}{02}}, \bibinfo{pages}{012}
  (\bibinfo{year}{2017}{\natexlab{a}}), \eprint{1609.08748}.

\bibitem[{\citenamefont{Cao et~al.}(2017{\natexlab{b}})\citenamefont{Cao,
  Zheng, Biesiada, Qi, Chen, and Zhu}}]{Cao:2017ivt}
\bibinfo{author}{\bibfnamefont{S.}~\bibnamefont{Cao}},
  \bibinfo{author}{\bibfnamefont{X.}~\bibnamefont{Zheng}},
  \bibinfo{author}{\bibfnamefont{M.}~\bibnamefont{Biesiada}},
  \bibinfo{author}{\bibfnamefont{J.}~\bibnamefont{Qi}},
  \bibinfo{author}{\bibfnamefont{Y.}~\bibnamefont{Chen}}, \bibnamefont{and}
  \bibinfo{author}{\bibfnamefont{Z.-H.} \bibnamefont{Zhu}},
  \bibinfo{journal}{Astron. Astrophys.} \textbf{\bibinfo{volume}{606}},
  \bibinfo{pages}{A15} (\bibinfo{year}{2017}{\natexlab{b}}),
  \eprint{1708.08635}.

\bibitem[{\citenamefont{Cao et~al.}(2018{\natexlab{a}})\citenamefont{Cao,
  Biesiada, Qi, Pan, Zheng, Xu, Ji, and Zhu}}]{Cao:2017abj}
\bibinfo{author}{\bibfnamefont{S.}~\bibnamefont{Cao}},
  \bibinfo{author}{\bibfnamefont{M.}~\bibnamefont{Biesiada}},
  \bibinfo{author}{\bibfnamefont{J.}~\bibnamefont{Qi}},
  \bibinfo{author}{\bibfnamefont{Y.}~\bibnamefont{Pan}},
  \bibinfo{author}{\bibfnamefont{X.}~\bibnamefont{Zheng}},
  \bibinfo{author}{\bibfnamefont{T.}~\bibnamefont{Xu}},
  \bibinfo{author}{\bibfnamefont{X.}~\bibnamefont{Ji}}, \bibnamefont{and}
  \bibinfo{author}{\bibfnamefont{Z.-H.} \bibnamefont{Zhu}},
  \bibinfo{journal}{Eur. Phys. J. C} \textbf{\bibinfo{volume}{78}},
  \bibinfo{pages}{749} (\bibinfo{year}{2018}{\natexlab{a}}),
  \eprint{1708.08639}.

\bibitem[{\citenamefont{Cao et~al.}(2019{\natexlab{b}})\citenamefont{Cao, Qi,
  Biesiada, Zheng, Xu, Pan, and Zhu}}]{cao2019milliarcsecond}
\bibinfo{author}{\bibfnamefont{S.}~\bibnamefont{Cao}},
  \bibinfo{author}{\bibfnamefont{J.}~\bibnamefont{Qi}},
  \bibinfo{author}{\bibfnamefont{M.}~\bibnamefont{Biesiada}},
  \bibinfo{author}{\bibfnamefont{X.}~\bibnamefont{Zheng}},
  \bibinfo{author}{\bibfnamefont{T.}~\bibnamefont{Xu}},
  \bibinfo{author}{\bibfnamefont{Y.}~\bibnamefont{Pan}}, \bibnamefont{and}
  \bibinfo{author}{\bibfnamefont{Z.}~\bibnamefont{Zhu}},
  \bibinfo{journal}{Physics of the Dark Universe}
  \textbf{\bibinfo{volume}{24}}, \bibinfo{pages}{100274}
  (\bibinfo{year}{2019}{\natexlab{b}}).

\bibitem[{\citenamefont{Foreman-Mackey
  et~al.}(2013)\citenamefont{Foreman-Mackey, Hogg, Lang, and
  Goodman}}]{Foreman_Mackey_2013}
\bibinfo{author}{\bibfnamefont{D.}~\bibnamefont{Foreman-Mackey}},
  \bibinfo{author}{\bibfnamefont{D.~W.} \bibnamefont{Hogg}},
  \bibinfo{author}{\bibfnamefont{D.}~\bibnamefont{Lang}}, \bibnamefont{and}
  \bibinfo{author}{\bibfnamefont{J.}~\bibnamefont{Goodman}},
  \bibinfo{journal}{Publications of the Astronomical Society of the Pacific}
  \textbf{\bibinfo{volume}{125}}, \bibinfo{pages}{306} (\bibinfo{year}{2013}).

\bibitem[{\citenamefont{Liu et~al.}(2020{\natexlab{a}})\citenamefont{Liu, Cao,
  Liu, Li, Geng, Lian, and Guo}}]{Liuyuting2020}
\bibinfo{author}{\bibfnamefont{Y.}~\bibnamefont{Liu}},
  \bibinfo{author}{\bibfnamefont{S.}~\bibnamefont{Cao}},
  \bibinfo{author}{\bibfnamefont{T.}~\bibnamefont{Liu}},
  \bibinfo{author}{\bibfnamefont{X.}~\bibnamefont{Li}},
  \bibinfo{author}{\bibfnamefont{S.}~\bibnamefont{Geng}},
  \bibinfo{author}{\bibfnamefont{Y.}~\bibnamefont{Lian}}, \bibnamefont{and}
  \bibinfo{author}{\bibfnamefont{W.}~\bibnamefont{Guo}}, \bibinfo{journal}{The
  Astrophysical Journal} \textbf{\bibinfo{volume}{901}}, \bibinfo{pages}{129}
  (\bibinfo{year}{2020}{\natexlab{a}}).

\bibitem[{\citenamefont{Yu and Wang}(2016)}]{Yu:2016gmd}
\bibinfo{author}{\bibfnamefont{H.}~\bibnamefont{Yu}} \bibnamefont{and}
  \bibinfo{author}{\bibfnamefont{F.}~\bibnamefont{Wang}},
  \bibinfo{journal}{Astrophys. J.} \textbf{\bibinfo{volume}{828}},
  \bibinfo{pages}{85} (\bibinfo{year}{2016}), \eprint{1605.02483}.

\bibitem[{\citenamefont{Wang et~al.}(2020{\natexlab{b}})\citenamefont{Wang, Ma,
  and Xia}}]{Wang:2020dbt}
\bibinfo{author}{\bibfnamefont{G.-J.} \bibnamefont{Wang}},
  \bibinfo{author}{\bibfnamefont{X.-J.} \bibnamefont{Ma}}, \bibnamefont{and}
  \bibinfo{author}{\bibfnamefont{J.-Q.} \bibnamefont{Xia}},
  \bibinfo{journal}{arXiv preprint arXiv:2004.13913}
  (\bibinfo{year}{2020}{\natexlab{b}}).

\bibitem[{\citenamefont{Zheng et~al.}(2017)\citenamefont{Zheng, Biesiada, Cao,
  Qi, and Zhu}}]{Zheng:2017asg}
\bibinfo{author}{\bibfnamefont{X.}~\bibnamefont{Zheng}},
  \bibinfo{author}{\bibfnamefont{M.}~\bibnamefont{Biesiada}},
  \bibinfo{author}{\bibfnamefont{S.}~\bibnamefont{Cao}},
  \bibinfo{author}{\bibfnamefont{J.}~\bibnamefont{Qi}}, \bibnamefont{and}
  \bibinfo{author}{\bibfnamefont{Z.-H.} \bibnamefont{Zhu}},
  \bibinfo{journal}{JCAP} \textbf{\bibinfo{volume}{10}}, \bibinfo{pages}{030}
  (\bibinfo{year}{2017}), \eprint{1705.06204}.

\bibitem[{\citenamefont{Qi et~al.}(2017)\citenamefont{Qi, Cao, Biesiada, Zheng,
  and Zhu}}]{Qi:2017xzl}
\bibinfo{author}{\bibfnamefont{J.-Z.} \bibnamefont{Qi}},
  \bibinfo{author}{\bibfnamefont{S.}~\bibnamefont{Cao}},
  \bibinfo{author}{\bibfnamefont{M.}~\bibnamefont{Biesiada}},
  \bibinfo{author}{\bibfnamefont{X.}~\bibnamefont{Zheng}}, \bibnamefont{and}
  \bibinfo{author}{\bibfnamefont{Z.-H.} \bibnamefont{Zhu}},
  \bibinfo{journal}{Eur. Phys. J. C} \textbf{\bibinfo{volume}{77}},
  \bibinfo{pages}{502} (\bibinfo{year}{2017}), \eprint{1708.08603}.

\bibitem[{\citenamefont{Liu et~al.}(2020{\natexlab{b}})\citenamefont{Liu, Cao,
  Biesiada, Liu, Geng, and Lian}}]{Liu:2020bac}
\bibinfo{author}{\bibfnamefont{T.}~\bibnamefont{Liu}},
  \bibinfo{author}{\bibfnamefont{S.}~\bibnamefont{Cao}},
  \bibinfo{author}{\bibfnamefont{M.}~\bibnamefont{Biesiada}},
  \bibinfo{author}{\bibfnamefont{Y.}~\bibnamefont{Liu}},
  \bibinfo{author}{\bibfnamefont{S.}~\bibnamefont{Geng}}, \bibnamefont{and}
  \bibinfo{author}{\bibfnamefont{Y.}~\bibnamefont{Lian}}, \bibinfo{journal}{The
  Astrophysical Journal} \textbf{\bibinfo{volume}{899}}, \bibinfo{pages}{71}
  (\bibinfo{year}{2020}{\natexlab{b}}).

\bibitem[{\citenamefont{Oguri and Marshall}(2010)}]{oguri2010gravitationally}
\bibinfo{author}{\bibfnamefont{M.}~\bibnamefont{Oguri}} \bibnamefont{and}
  \bibinfo{author}{\bibfnamefont{P.~J.} \bibnamefont{Marshall}},
  \bibinfo{journal}{Monthly Notices of the Royal Astronomical Society}
  \textbf{\bibinfo{volume}{405}}, \bibinfo{pages}{2579} (\bibinfo{year}{2010}).

\bibitem[{\citenamefont{Pereira et~al.}(2013)\citenamefont{Pereira, Thomas,
  Aldering, Antilogus, Baltay, Benitezherrera, Bongard, Buton, Canto,
  Cellierholzem et~al.}}]{pereira2013}
\bibinfo{author}{\bibfnamefont{R.}~\bibnamefont{Pereira}},
  \bibinfo{author}{\bibfnamefont{R.~C.} \bibnamefont{Thomas}},
  \bibinfo{author}{\bibfnamefont{G.}~\bibnamefont{Aldering}},
  \bibinfo{author}{\bibfnamefont{P.}~\bibnamefont{Antilogus}},
  \bibinfo{author}{\bibfnamefont{C.}~\bibnamefont{Baltay}},
  \bibinfo{author}{\bibfnamefont{S.}~\bibnamefont{Benitezherrera}},
  \bibinfo{author}{\bibfnamefont{S.}~\bibnamefont{Bongard}},
  \bibinfo{author}{\bibfnamefont{C.}~\bibnamefont{Buton}},
  \bibinfo{author}{\bibfnamefont{A.}~\bibnamefont{Canto}},
  \bibinfo{author}{\bibfnamefont{F.}~\bibnamefont{Cellierholzem}},
  \bibnamefont{et~al.}, \bibinfo{journal}{Astronomy and Astrophysics}
  \textbf{\bibinfo{volume}{554}}, \bibinfo{pages}{1} (\bibinfo{year}{2013}).

\bibitem[{\citenamefont{Goldstein and Nugent}(2016)}]{goldstein2016how}
\bibinfo{author}{\bibfnamefont{D.~A.} \bibnamefont{Goldstein}}
  \bibnamefont{and} \bibinfo{author}{\bibfnamefont{P.}~\bibnamefont{Nugent}},
  \bibinfo{journal}{The Astrophysical Journal} \textbf{\bibinfo{volume}{834}}
  (\bibinfo{year}{2016}).

\bibitem[{\citenamefont{Pushkarev and Kovalev}(2015)}]{PK15}
\bibinfo{author}{\bibfnamefont{A.}~\bibnamefont{Pushkarev}} \bibnamefont{and}
  \bibinfo{author}{\bibfnamefont{Y.}~\bibnamefont{Kovalev}},
  \bibinfo{journal}{Monthly Notices of the Royal Astronomical Society}
  \textbf{\bibinfo{volume}{452}}, \bibinfo{pages}{4274} (\bibinfo{year}{2015}).

\bibitem[{\citenamefont{Cao et~al.}(2018{\natexlab{b}})\citenamefont{Cao,
  Biesiada, Qi, Pan, Zheng, Xu, Ji, and Zhu}}]{cao2018cosmological}
\bibinfo{author}{\bibfnamefont{S.}~\bibnamefont{Cao}},
  \bibinfo{author}{\bibfnamefont{M.}~\bibnamefont{Biesiada}},
  \bibinfo{author}{\bibfnamefont{J.}~\bibnamefont{Qi}},
  \bibinfo{author}{\bibfnamefont{Y.}~\bibnamefont{Pan}},
  \bibinfo{author}{\bibfnamefont{X.}~\bibnamefont{Zheng}},
  \bibinfo{author}{\bibfnamefont{T.}~\bibnamefont{Xu}},
  \bibinfo{author}{\bibfnamefont{X.}~\bibnamefont{Ji}}, \bibnamefont{and}
  \bibinfo{author}{\bibfnamefont{Z.}~\bibnamefont{Zhu}},
  \bibinfo{journal}{European Physical Journal C} \textbf{\bibinfo{volume}{78}},
  \bibinfo{pages}{749} (\bibinfo{year}{2018}{\natexlab{b}}).

\bibitem[{\citenamefont{Cao et~al.}(2020)\citenamefont{Cao, Qi, Biesiada, Liu,
  and Zhu}}]{cao2020precise}
\bibinfo{author}{\bibfnamefont{S.}~\bibnamefont{Cao}},
  \bibinfo{author}{\bibfnamefont{J.}~\bibnamefont{Qi}},
  \bibinfo{author}{\bibfnamefont{M.}~\bibnamefont{Biesiada}},
  \bibinfo{author}{\bibfnamefont{T.}~\bibnamefont{Liu}}, \bibnamefont{and}
  \bibinfo{author}{\bibfnamefont{Z.}~\bibnamefont{Zhu}}, \bibinfo{journal}{The
  Astrophysical Journal} \textbf{\bibinfo{volume}{888}} (\bibinfo{year}{2020}).

\end{thebibliography}

\end{document}